\documentstyle[aps,preprint,epsfig,tighten]{revtex}
\newcommand{\BE}{\begin{equation}}
\newcommand{\EE}{\end{equation}}
\newcommand{\BEA}{\begin{eqnarray}}
\newcommand{\EEA}{\end{eqnarray}}
\newcommand{\llangle}{\langle\!\langle}
\newcommand{\rrangle}{\rangle\!\rangle}
\begin{document}
\draft
\title{
Two- and Three-Pion Interferometry for a Nonchaotic Source in
Relativistic Nuclear Collisions
}
\author{Hiroki Nakamura$^{(1)}$ and Ryoichi Seki$^{(2,3)}$}
\address{
${}^{(1)}$ Department of Physics,
Waseda University, Tokyo 169-8555, Japan \\
${}^{(2)}$ Department of Physics, California State University,
Northridge, CA 91330 \\
${}^{(3)}$ W.~K.~Kellogg Radiation Laboratory, 106-38, 
California Institute of Technology, \\
Pasadena, CA 91125} 
\date{\today}
\maketitle
\begin{abstract}
Two- and three-pion correlation functions are investigated
for a source that is not fully chaotic.
Various models are examined to describe the source.
The chaoticity and weight factor are evaluated in each model
as measures of the strength of correlations
and compared to experimental results.
A new measure of three-pion correlation is also suggested.
\end{abstract}

\section{Introduction}

Two-pion correlations obtained in relativistic heavy-ion collisions
have been used to extract the size and shape of
the pion-emitting source, based on the Hanbury-Brown Twiss (HBT) effect.
Correlations are not, of course, limited to those of two pions, but 
can also be of multi-pions.  Though multi-pion correlations are unavoidably
complicated, they contain new information not available from the two-pion
correlations.  For example, when final-state interactions of the emitted
pions are neglected, the two-pion correlations for a chaotic source
depend only on the magnitude
of the Fourier transform of the pion-source function, but the multi-pion
correlations depend also on its phase\cite{hz,ns}.  The magnitude
is an even function of the relative momentum of the emitted pions, while
the phase is an odd function.  One thus hopes to extract new information
about the source from the multi-pion correlations.

In a previous article\cite{ns}, we made a detailed investigation
of three-pion correlations, 
the simplest multi-pion correlations after the two-pion correlations,
over a wide range of kinematics, to investigate the feasibility of
extracting new information about the source through the phase
in the case of a chaotic source.  We find that extracting new information
would be rather difficult in practice because the multiplicative factor
of the phase function becomes small in the region where the phase varies
appreciably.

In experiments, the two-pion correlations at the zero relative momentum
is observed to be less than two.  After the final-state interactions
are removed, it must be two in the case of a chaotic source.  The
measurement suggests that the source is not completely chaotic.
Recently, a measurement of the three-pion correlations has been reported,
showing that the strength of the three-pion correlations are also
less than expected in the case of a chaotic source.

In this paper we investigate the two- and three-pion correlations
for a source not completely chaotic.  We introduce models of the source
that are various mixtures of coherent and chaotic sources.  The models
also include one with a novel structure, a mixture of multiple coherent
sources and a chaotic source.  These models describe different dynamics
generating pion emissions, though we do not pursue the identification of 
dynamics corresponding to each model.
 
A mixture of a coherent source and a chaotic source is a model that has
often appeared in the literature\cite{gkw,biyajima,hz} under the name 
``partially coherent source.''   We find that this popular model poorly
reproduces the recent data of the two- and three-pion correlations\cite{na44}.
A model consisting of multi-coherent sources and one chaotic source 
appears to yield a good agreement.

It is both theoretical and experimental practice to calculate the so-called
weight factor from the three-pion correlations at the vanishing relative
momenta.  The weight factor is usually considered to represent the strength
of the genuine three-pion correlations. When we examine more complicated
models than the popular partially coherent model, we find that the weight
factor no longer describes the genuine strength.  The expression that
yields the genuine strength depends on the structure of the source, and
there is no universal expression as such.  Nevertheless, we propose a modified
expression for the weight factor that has a wider validity.

In Sec. \ref{sec2}, we define pion spectra, correlation functions,
and measures of two- and three-pion correlation functions, chaoticity, and
weight factor, respectively. In Sec. III, various nonchaotic models of
the pion-emitting source are introduced, and chaoticities and weight factors
are obtained for them.  Section \ref{sec3} presents discussions and a summary:
a new expression of the weight factor is introduced and discussed, 
and chaoticities and weight factors are compared 
with the recent experimental data.
Appendix \ref{appA} gives a derivation of correlations for multi-coherent
sources. In Appendix \ref{appB}, the new weight factor is derived.

\section{Correlation functions, chaoticity, and weight factor}
\label{sec2}

For the sake of clarity in the sections to follow, we define correlation
functions and their measures, chaoticity, and weight factor.  Note that our
definition is standard.

We first write the basic quantities, pion spectra, in second-quantization
form, as follows:
\begin{eqnarray}
\label{eq:w1}
W_1(p) &=& \langle a_p^\dagger a_p\rangle,\\
W_2(p_1,p_2) &=&  \langle a_{p_1}^\dagger a_{p_2}^\dagger
            {a_{p_1}}{a_{p_2}}\rangle, \\
W_3(p_1,p_2,p_3) &=&  \langle a_{p_1}^\dagger
             a_{p_2}^\dagger
             a_{p_3}^\dagger{a_{p_1}}
            {a_{p_2}}{a_{p_3}}\rangle.
\label{eq:w3}
\end{eqnarray}
Though these expressions are simple and reasonable, the explicit definition of
$\langle ...\rangle$ is a complicated issue.
$\langle ...\rangle$ represents a quantum statistical average and is
formally written as $\langle\psi|...|\psi\rangle$ in terms of the quantum
state, $|\psi\rangle$, or ${\mathrm{Tr}}\{\hat{\rho}...\}$ in terms of the 
the density matrix, $\hat{\rho}$.  In this work, we introduce models
to represent various underlying dynamics of the pion emission.
Note that the momenta above and hereafter are on-shell,
e.g., $p^0 =\sqrt{{\mathbf{p}}^2+m^2}$ for $p$.

In terms of the spectra, we define two- and three-pion 
correlation functions in the usual
way\cite{gkw},
\BEA
\label{eq:c2def}
C_2(p_1,p_2)
    &=& \frac{\langle n \rangle^2}{\langle n(n-1) \rangle}
        \frac{W_2(p_1,p_2)}{W_1(p_1)W_1(p_2)},\\
\label{eq:c3def}
C_3(p_1,p_2,p_3) &=&
  \frac{\langle n \rangle^3}{\langle n(n-1)(n-2) \rangle}
     \frac{W_3(p_1,p_2,p_3)}{W_1(p_1)W_1(p_2)W_1(p_3)}.
\EEA
Here, the normalizations are introduced in order to take account of
multiplicity fluctuation, with the following definitions:
\BEA
\label{eq:nave}
\langle n \rangle &=& \int d^3p W_1(p), \\
\langle n(n-1) \rangle &=& \int d^3p_1d^3p_2 W_2(p_1,p_2), \\
\langle n(n-1)(n-2) \rangle &=& 
\int d^3p_1d^3p_2d^3p_3 W_2(p_1,p_2,p_3).
\label{eq:naveend}
\EEA
In the usual models, the above normalizations yield $C_2$ and
$C_3 \rightarrow 1$ as the relative momenta approach infinity.
In more complicated models, such as those that possess the particle-number
fluctuation per mode, however, the normalizations of Eqs. (\ref{eq:c2def})
and (\ref{eq:c3def}) yield an asymptotic value different from unity.
We will discuss this point fully in Sec. \ref{sec:mcha}.

When the HBT effect does not appear, correlation functions are independent
of relative momenta.  When the HBT effect occurs, however, the correlation
functions are no longer constant, but are dependent on relative momenta.
As is well known, the size and shape of a pion-emitting source are extracted
from the relative-momentum dependence of the correlations through the HBT
effect.  The methods of extracting the size and shape of the source has been
discussed numerous times, and we will not go into the issues here.

We define the chaoticity, $\lambda(p)$, and the weight factor, $\omega(p)$,
which are usually considered to represent measures of the strength 
of the two- and three-pion correlations,
respectively\cite{hz,na44}:
\begin{eqnarray}
\label{eq:chaoticity}
\lambda(p) &=& {C}_2(p,p) - 1,\\
\label{eq:wf}
\omega(p) &=& \left.\frac{{C}_3(p_1,p_2,p_2)-1
   -(C_2(p_1,p_2)-1)
   -(C_2(p_2,p_3)-1)
   -(C_2(p_3,p_1)-1)
}
  {2\sqrt{(C_2(p_1,p_2)-1)(C_2(p_2,p_3)-1)(C_2(p_3,p_1)-1)}}
\right|_{p_1=p_2=p_3=p}. \nonumber \\
\end{eqnarray}
Note that we define the weight factor explicitly at $p_1=p_2=p_3$, but
that it has been extracted from measurements over small relative
momenta\cite{na44}.  The two methods show practically no difference
since the weight factor is expected to be a slowly varying function
of the relative momenta\cite{hz,ns}.

The weight factor, however, is not always the measure of the strength
of the genuine three-pion correlations. 
Generally, the three-pion correlation functions are related to the
two-pion correlation functions as
\BEA
\label{eq:c3}
{C}_3(p_1,p_2,p_3) &=& 1 + a ({C}_2(p_1,p_2)
+ {C}_2(p_2,p_3)+ {C}_2(p_3,p_1) -3)
 \nonumber \\
&&+ (\mbox{genuine three-pion correlation}) 
\nonumber \\&&
+ (\mbox{other two-pion
correlation}),
\EEA
where the coefficient, $a$, is not always unity even at $p_1 = p_2 = p_3$.
Consequently, Eq. (\ref{eq:wf}) does not always remove the linear dependence
of $C_2$'s from $C_3$. In the cases of a chaotic source and a partially coherent
source, $a$ is unity, but for sources of more complicated structure, it is
not.  Furthermore, $C_3$ generally depends on the two-pion correlations not
only as the linear $C_2$'s but also in more complicated ways as a function of
two momenta.  We will discuss these points fully in the following section.

The relation between $C_3$ and $C_2$ depends on dynamics and is
model-dependent in practice.  The extraction of the strength of the genuine 
three-pion correlations thus requires explicit knowledge of the dynamics.
Accordingly, in this paper we adopt the conventional approach of using
the weight factor defined as Eq. (\ref{eq:wf}), for numerical results
as a convenient means for making comparisons among various models.

\section{Different models of the source}

\subsection{Coherent Source and Chaotic Source}

The spectra are often written by $c$-number source current\cite{gkw,apw},
yielding a chaotic source and a coherent source as the two extreme cases.
The chaotic source shows the HBT effect with both the chaoticity and the
weight factor as unity.  The correlation functions for a coherent source
are independent of the relative momenta, and the chaoticity for it is zero.

The difference between the out-state and in-state annihilation operators of
an emitted pion defines the source current as
\begin{eqnarray}
a_{\rm{out}}({\bf{p}}) &=& a_{\rm{in}}({\bf{p}})
 + i \int d^4x\frac{1}{ \sqrt{(2\pi)^3\cdot 2p^0}} J(x) e^{-i p\cdot x}
\nonumber \\
&=&  a_{\rm{in}}({\bf{p}})
 + i \frac{1}{ \sqrt{(2\pi)^3\cdot 2p^0}} J(p),
\end{eqnarray}
where $J(x)$ and $J(p)$ are the source current and its Fourier transform,
respectively. 
The average that appears in Eqs. (\ref{eq:w1})-(\ref{eq:w3})
 can be written as
\BE
\langle \cdots \rangle
 = \int {\cal D}J^*(p){\cal D}J(p){\cal P}[J^*(p),J(p)]
 \langle 0_{in}|\cdots|0_{in}\rangle,
\EE
where $|0_{in}\rangle$ is the in-state vacuum and ${\cal P}[J^*(p),J(p)]$
is the distribution functional of $J(p)$, which has a statistical
fluctuation.
The pion spectra are obtained in this case  from
Eqs. (\ref{eq:w1})-(\ref{eq:w3}):
\begin{eqnarray}
\label{eq:onesp}
W_1(p) &=& \frac{1}{ (2\pi)^3\cdot 2p^0} \langle |J(p)|^2\rangle_J,\\
\label{eq:twosp}
W_2(p_1,p_2) 
&=& \frac{1}{(2\pi)^6\cdot 2{p^0_1}\cdot 2{p^0_2}}
   \langle |J(p_1)|^2 |J(p_2)|^2\rangle_J, \\
\label{eq:thrsp}
W_3(p_1,p_2,p_3) 
&=& \frac{1}{(2\pi)^9\cdot 2{p^0_1}\cdot 2{p^0_2}\cdot 2{p^0_3}}
   \langle |J(p_1)|^2 |J(p_2)|^2 |J(p_3)|^2\rangle_J,
\end{eqnarray}
where $\langle \cdots \rangle_J = \int {\cal D}J^*(p){\cal D}J(p)
{\cal P}[J^*(p),J(p)]\cdots$.
Hereafter, we do not explicitly show the subscript $J$ of the angle bracket.
If the phase of the source current is random, we call the source {\it chaotic}.
In this case, ${\cal P}[J^*(p),J(p)]$ have a Gaussian form\cite{apw} and
higher-order moments of $J(p)$ are represented by the second-order moment
such as
\begin{eqnarray}
\lefteqn{
\langle J^*(p_1) J^* (p_2) J(q_1) J(q_2) \rangle
 =} \nonumber \\
&& \langle J^*(p_1) J(q_1) \rangle
   \langle J^*(p_2) J(q_2) \rangle
 + \langle J^*(p_1) J(q_2) \rangle
   \langle J^*(p_2) J(q_1) \rangle.
\end{eqnarray}
For a chaotic source, the pion spectra are written as
\begin{eqnarray}
W_1(p_1) &=& F_{11}, \\
W_2(p_1,p_2) &=& F_{11} F_{22} + {F_{12}}^2, \\
W_3(p_1,p_2,p_3) &=& F_{11} F_{22} F_{33} 
+\sum_{(a,b,c)}F_{aa}{F_{bc}}^2
+2F_{12}F_{23}F_{31}\cos(\Phi_{12}+\Phi_{23}+\Phi_{31}),
\end{eqnarray}
where the amplitude, $F_{ab}$, and the phase, $\Phi_{ab}$ 
are defined as\cite{hz}
\begin{equation}
\label{eq:chaF}
 F_{ab}\exp(i\Phi_{ab}) \equiv 
   \frac{1}{(2\pi)^3\sqrt{2{p^0_a}\cdot 2{p^0_b}}} 
   \langle J^{*}(p_a)J(p_b) \rangle. 
\end{equation}
$\sum_{(a,b,c)}$ implies a sum over $(a,b,c)=(1,2,3),(2,3,1),(3,1,2)$.
The correlation functions are then
\begin{eqnarray}
C_2(p_1,p_2) &=& 1+\frac{F^{2}_{12}}{F_{11}F_{22}}, \\
C_3(p_1,p_2,p_3)&=& 1
+ \sum_{(a,b)}\frac{{F_{ab}}^2}{F_{aa}F_{bb}}
+ 2 \frac{F_{12}F_{23}F_{31}}{F_{11}F_{22}F_{33}}
  \cos(\Phi_{12}+\Phi_{23}+\Phi_{31}),
\end{eqnarray}
where $\sum_{(a,b)}$ is over $(a,b)=(1,2),(2,3),(3,1)$.
$F_{ab}$ and $\Phi_{ab}$ contain the information about the size and shape
of the pion-emitting source, due to the HBT effect.
The chaoticity and weight factor for a chaotic source always achieve unity.

When the source current has no randomness, the source is called
{\it coherent}.  In this case, the $n$-pion spectra are expressed as 
\begin{equation}
W_n(p_1,...,p_n) = \prod_{l=1}^nW_1(p_l).
\end{equation}
The HBT effect does not appear in this case. The correlation functions
achieve unity, and the chaoticity vanishes.

\subsection{Partially Coherent Source}
\label{sbsec:pcs}

In nuclear collisions, one may expect to involve the mixture of a coherent
source and a chaotic source, as suggested by the observation that the
chaoticities are often smaller than unity. Here, we sketch the case of the
partially coherent source, mostly following Ref.\cite{hz}.

\begin{figure}
\begin{center}
\epsfig{file=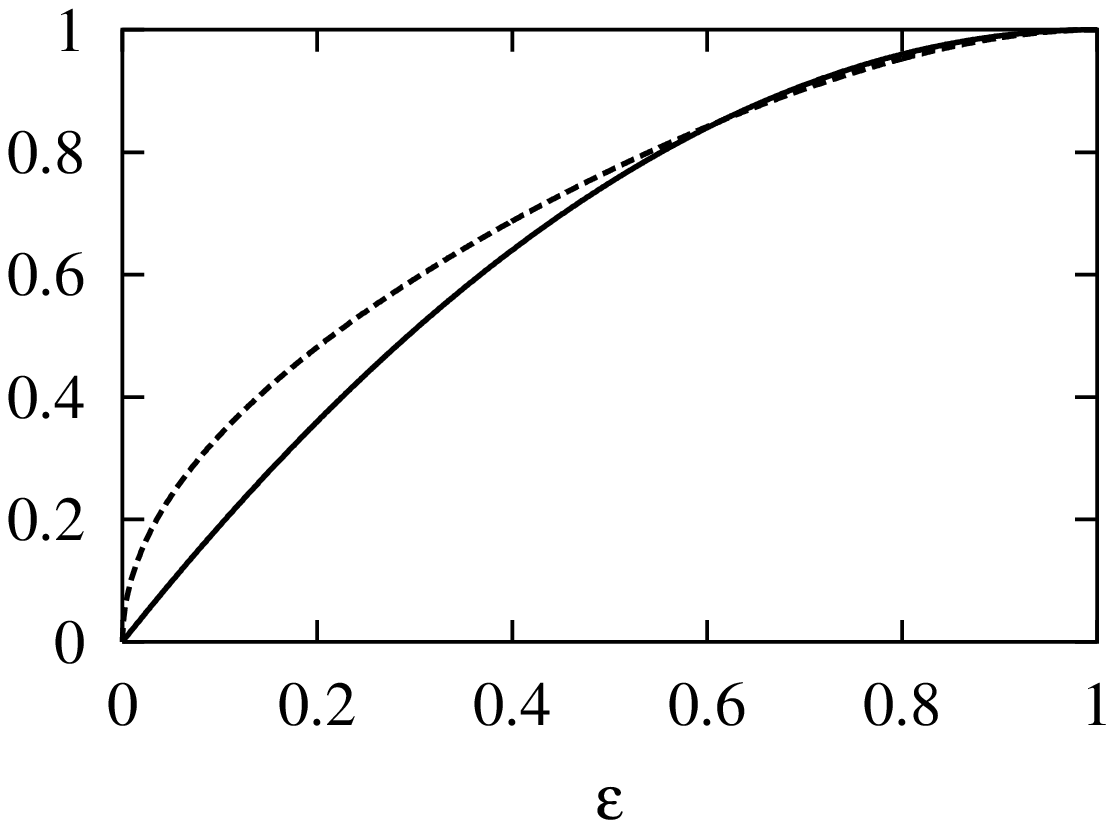 }%
\caption{
Chaoticity and weight factor as functions of $\epsilon$ 
in the partially coherent model. 
Solid and dashed 
lines stand for the chaoticity and weight factor, respectively.}
\label{fig:parcoh}
\end{center}
\end{figure}

When the pion-emitting source is a mixture of a chaotic source and a
coherent source, the source current is written as a sum of both currents,
$J(x)=J_{cha}(x)+J_{coh}(x)$\cite{hz,gkw}.
The sources are not correlated with each other,
$\langle J^*_{cha}(p)J_{coh}(q)\rangle$=0.
The one-pion spectrum and correlation functions for this source
are obtained as
\begin{eqnarray}
W_1(p_1) &=& f_{11}+F_{11}, \\
C_2(p_1,p_2) &=&
   1+\frac{{F_{12}}^2+2f_{12}F_{12}\cos(\Phi_{12}-\phi_{12})}
      {(f_{11}+F_{11})(f_{22}+F_{22})}, \\
C_3(p_1,p_2,p_3)&=& 1
  + \sum_{(a,b)} \frac{{F_{ab}}^2+2f_{ab}F_{ab}\cos(\Phi_{ab}-\phi_{ab})}
      {(f_{aa}+F_{aa})(f_{bb}+F_{bb})}
\nonumber \\ &&
 + 2\frac{1}{(f_{11}+F_{11})(f_{22}+F_{22})(f_{33}+F_{33})}
     \bigg\{{F_{12}F_{23}F_{31}}
       \cos(\Phi_{12}+\Phi_{23}+\Phi_{31})     
\nonumber \\ && 
    +\sum_{(a,b,c)} \left[
         {f_{ab}F_{bc}F_{ca}} \cos(\phi_{ab}+\Phi_{bc}+\Phi_{ca})
     \right]
    \bigg\},
\end{eqnarray}
where
\begin{equation}
 f_{ab}\exp(i\phi_{ab}) \equiv
   \frac{1}{(2\pi)^3\sqrt{2{p^0_a}\cdot 2{p^0_b}}}
   J_{coh}^{*}(p_a)J_{coh}(p_b),
\end{equation}
and $F_{ab}$ and $\Phi_{ab}$ are the same as those for a chaotic source,
Eq. (\ref{eq:chaF}).
The chaoticity and weight factor are evaluated as
\begin{eqnarray}
\label{eq:chapcs}
\lambda(p) &=& \epsilon(p)(2-\epsilon(p)), \\
\omega(p) &=& 
\sqrt{\epsilon(p)}\frac{3-2\epsilon(p)}{(2-\epsilon(p))^{3/2}},
\end{eqnarray}
where $\epsilon(p)$ is the fractional parameter of the coherent source:
\begin{equation}
\epsilon(p_1) = \frac{F_{11}}{f_{11}+F_{11}}.
\end{equation}
The source becomes chaotic for $\epsilon(p)=1$, while it becomes
coherent for $\epsilon(p)=0$.
Figure \ref{fig:parcoh} shows the chaoticity and weight factor
as functions of $\epsilon(p)$, and in Fig. \ref{fig:nondcc}
the weight factor is shown as a function of chaoticity,
varying as $\epsilon(p)$.
The weight factor can vary from 0 to 1.  In this case, 
the term $-3\lambda$, in Eq. (\ref{eq:wf}), completely removes 
the two-pion correlations.

\subsection{Multi-Coherent Sources}

We consider the mixture of a small number of coherent sources.  This model
differs from the case of multiple coherent sources that was previously examined
in \cite{gkw} and, to the best of our knowledge, this is the first time
that this model has been explicitly discussed.
We assume here that the pion-emitting source is made
of $N$ coherent sources that are not coherent with 
each other and {\it appear to be obeying the Poisson distribution.}
Note that the model considered in \cite{gkw} is a mixture of multiple
coherent sources that are randomly distributed, and that the model 
was introduced as
a description of a chaotic source when the number of the coherent
sources becomes large.  In a sense our model here is the opposite limit
of the small number of coherent sources in the model considered
in \cite{gkw}.  In the following, we present merely the final expressions,
leaving their derivations for Appendix A. 

The source current can be written as
\begin{equation}
J(x)= \sum_{n=1}^N j(x-X_n)e^{-i\theta_n},
\end{equation}
where $j(x-X_n)$ is the n-th coherent source current, 
located at $X_n$ with the random phase $\theta_n$.
Each coherent source is assumed to be expressed by 
the same $j(x)$ but to be located at a different position.
There are $N$ coherent sources, obeying the Poisson distribution.
Note that the distribution must be renormalized in order to exclude
the no-source event that is not observed.

The one-pion spectrum and normalized correlation functions 
are obtained as 
\begin{eqnarray}
W(p_1) &=& a_1 \frac{\alpha}{1-e^{-\alpha}}, \\
\label{eq:c2s}
{C}_2(p_1,p_2) &=& 1+\frac{\alpha}{\alpha+1} |\rho_{12}|^2, \\
\label{eq:c3s}
{C}_3(p_1,p_2,p_3) &=& 1+\frac{\alpha(\alpha+2)}{\alpha^2+3\alpha+1}
  \sum_{(a,b)}|\rho_{ab}|^2
 + \frac{\alpha^2}{\alpha^2+3\alpha+1}
     \cdot 2{\mathrm{Re}}\left(\rho_{12}\rho_{23}\rho_{31}\right),
\end{eqnarray}
where $\rho_{ij}$ is the Fourier transform of the spatial distribution 
of coherent sources, and $\alpha$ is a parameter of the Poisson
distribution. The mean number of coherent sources is
$\alpha/(1-\exp(-\alpha))$.
In this case, the subtraction $-3\lambda$, in Eq. (\ref{eq:wf}),
does not remove the two-pion correlations completely,
because the coefficient of $|\rho_{12}|^2$ in Eq. (\ref{eq:c2s})
is different from that in Eq. (\ref{eq:c3s}).
The chaoticity and weight factor are derived as
\begin{eqnarray}
{\lambda} &=& \frac{\alpha}{\alpha+1},\\
{\omega} &=&
  \frac{1}{2}\frac{2\alpha^2+2\alpha+3}{\alpha^2+3\alpha+1}
  \sqrt{\frac{\alpha+1}{\alpha}}.
\end{eqnarray}

\begin{figure}
\begin{center}
\epsfig{file=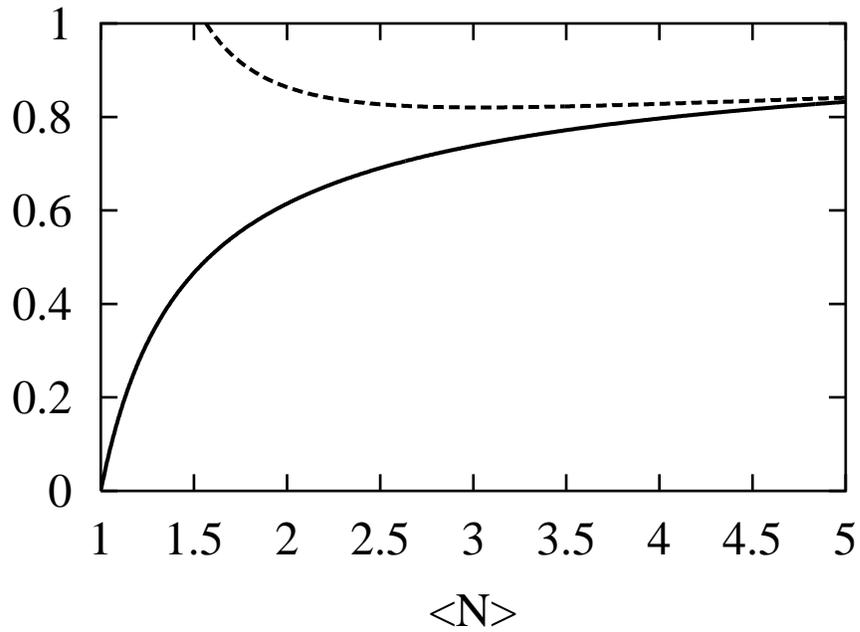}%
\caption{
Chaoticity and weight factor as functions of the mean number of
coherent sources  
in the model of multi-coherent sources. 
Solid and dashed 
lines stand for the chaoticity and weight factor, respectively.}
\label{fig:manycoh}
\end{center}
\end{figure}

In Fig. \ref{fig:manycoh}, the chaoticity and weight factor
are illustrated as functions of the mean number of sources,
 and Fig. \ref{fig:nondcc}
shows the weight factor as a function of the chaoticity.
The chaoticity varies from 0 to 1, because this source becomes coherent
at $\alpha=0$ and chaotic at $\alpha\to\infty$.
The weight factor diverges at $\alpha\to0$. The reason for this
is the failure of the subtraction $-3\lambda$, in Eq. (\ref{eq:wf}),
and the divergence is not caused by the genuine three-pion correlation.

\subsection{Multi-Coherent Sources and One Chaotic Source}
\label{sec:mcha}

We now consider the source in which the multi-coherent sources of the
previous subsection are mixed with a chaotic source.
The multi-coherent sources are not coherent with each other, as before.
The source current is written as
\BE
J(x) = \sum_{n=1}^Nj(x-X_n)e^{-i\theta_n} + J_{cha}(x),
\EE
where $j(x)$ and $J_{cha}(x)$ are a coherent source current and the
chaotic source current, respectively. 
The $n$-th coherent source is located at $X_n$ and is distributed
with $\rho(X_n)$.  
The phase, $\theta_n$, is randomly distributed between 0 and $2\pi$.
There are $N$ coherent sources, obeying the Poisson distribution,
Eq. (\ref{eq:usupoi}).

Pion spectra come out to be 
\BEA
W_1(p_1)&=& \alpha h_{11}+ F_{11}, \\
W_2(p_1,p_2) &=& W_1(p_1)W_1(p_2)+ \alpha h_{11}h_{22}
  +\left|\alpha h_{12}e^{i\psi_{12}}+F_{12}e^{i\Phi_{12}}\right|^2, \\
W_3(p_1,p_2,p_3) &=& W_1(p_1)W_1(p_2)W_1(p_3)+
 (3\alpha+1)\alpha h_{11}h_{22}h_{33}+\alpha\sum_{(a,b,c)}h_{aa}h_{bb}F_{cc}
\nonumber \\ &&
+\sum_{(a,b,c)}W_1(p_a)
   \left|\alpha h_{bc}e^{i\psi_{bc}}+F_{bc}e^{i\Phi_{bc}}\right|^2
\nonumber \\ &&
+ 2 \sum_{(a,b,c)}h_{aa}\left(\alpha^2 |h_{bc}|^2+\alpha h_{bc}F_{bc}\cos(\Phi_{bc}-\psi_{bc})\right)
\nonumber \\ &&
+2 {\mathrm{Re}}\left\{(\alpha h_{12}e^{i\psi_{12}}+F_{12}e^{i\Phi_{12}})
(\alpha h_{23}e^{i\psi_{23}}+F_{23}e^{i\Phi_{23}})
(\alpha h_{31}e^{i\psi_{31}}+F_{31}e^{i\Phi_{31}})
\right\}, \nonumber \\
\EEA
where $\alpha$ is the parameter in the Poisson distribution, and 
\BE
  h_{ij}e^{i\psi_{ij}}= 
  \frac{j^*(p_i)j(p_j)}{(2\pi)^3\sqrt{2p^0_1\cdot 2p_2^0}}\rho_{ij}.
\EE
For $|p_1-p_2|\to\infty$, we have
\BEA
\frac{W_2(p_1,p_2)}{W_1(p_1)W_1(p_2)}
&\to&
1+\frac{1}{\alpha}(1-\epsilon(p_1))(1-\epsilon(p_2)),
\EEA 
where the fractional parameter of the chaotic source, $\epsilon$, is
defined as
\BE
\epsilon(p_1) = \frac{F_{11}}{\alpha h_{11}+F_{11}}.
\EE
The normalization factor of $C_2(p_1,p_2)$, shown in Eq. (\ref{eq:c2def}),
does not yield the proper asymptotic value of unity as the relative momentum
approaches infinity.  Generally $\epsilon$ depends on the momentum, and
the asymptotic value of $C_2$ thus depends on the two momenta separately.
If we assume that $\epsilon$ is independent of the momentum,
we find that $C_2$ is normalized properly by the use of 
Eqs. (\ref{eq:c2def}) and (\ref{eq:c3def}).
In this case, the correlation functions are given by 
\BEA
C_2(p_1,p_2) &=& 1+
\frac{\left|\alpha h_{12}e^{i\psi_{12}}+F_{12}e^{i\Phi_{12}}\right|^2 }
{W_1(p_1)W_1(p_2)\left(1+\frac{1}{\alpha}(1-\epsilon)^2\right)}, \\
C_3(p_1,p_2,p_3) &=& 1+
\frac{1}{1+\frac{3\alpha+1}{\alpha^2}(1-\epsilon)^3
   +\frac{3}{\alpha}(1-\epsilon)^2\epsilon}
 \left[
   \sum_{(a,b)}
  \frac{\left|\alpha h_{ab}e^{i\psi_{ab}}+F_{ab}e^{i\Phi_{ab}}\right|^2 }
   {W_1(p_a)W_1(p_b)}
\right.\nonumber \\ &&
   +
  \frac{
    2 {\mathrm{Re}}\left\{(\alpha h_{12}e^{i\psi_{12}}+F_{12}e^{i\Phi_{12}})
      (\alpha h_{23}e^{i\psi_{23}}+F_{23}e^{i\Phi_{23}})
      (\alpha h_{31}e^{i\psi_{31}}+F_{31}e^{i\Phi_{31}})
     \right\}
   }{W_1(p_1)W_1(p_2)W_1(p_3)}
\nonumber \\ && \left.
 + 2\sum_{(a,b)}\left\{
    \frac{\alpha(1-\epsilon)h_{ab}^2}{W_1(p_a)W_1(p_b)}
    + \frac{(1-\epsilon) h_{ab}F_{ab}\cos(\Phi_{ab}-\psi_{ab})}
       { W_1(p_a)W_1(p_b)}
   \right\}
  \right].
\EEA
Comparison of the preceding $C_2$ and the first term in the square bracket in
$C_3$ shows that the two-pion correlations are not to be removed completely, as
$a \neq 1$ in Eq. (\ref{eq:c3}) in this case.  The conventional weight factor
thus no longer represents the strength of the genuine three-pion correlations.
Furthermore, the last sum in the square bracket of the above $C_3$
represents the effects of the two-pion correlation (because it depends on a pair
of the momenta), corresponding to the ``other two-pion correlation''
in Eq. (\ref{eq:c3}).

\begin{figure}
\begin{center}
\epsfig{file=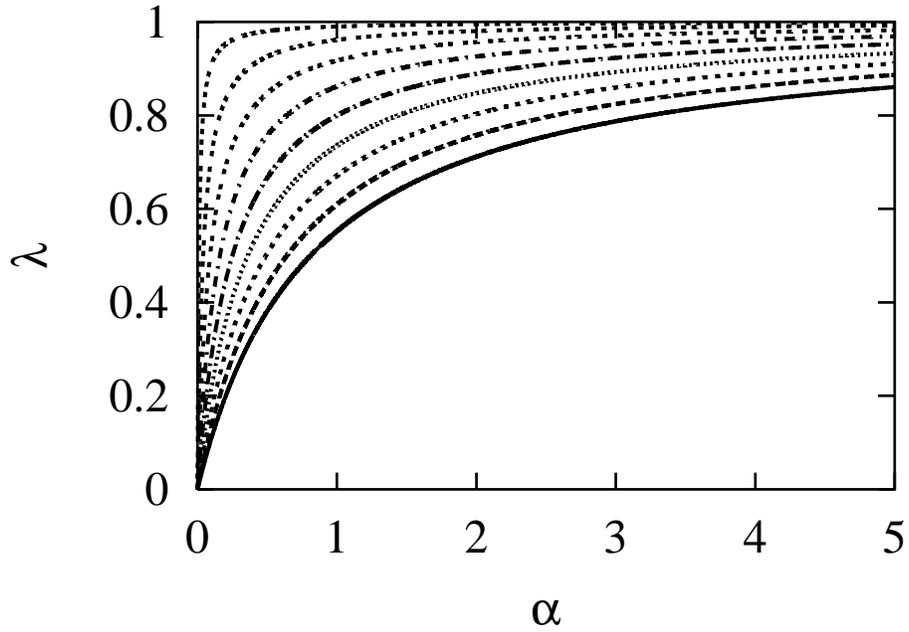}%
\caption{Chaoticity as a function of $\alpha$ in the model  of
multi-coherent sources and one chaotic source. The lines from down to up
correspond to  $\epsilon$ varying from 0.1 to 0.9 with the step 0.1. }
\label{fig:mchal}
\end{center}
\end{figure}

\begin{figure}
\begin{center}
\epsfig{file=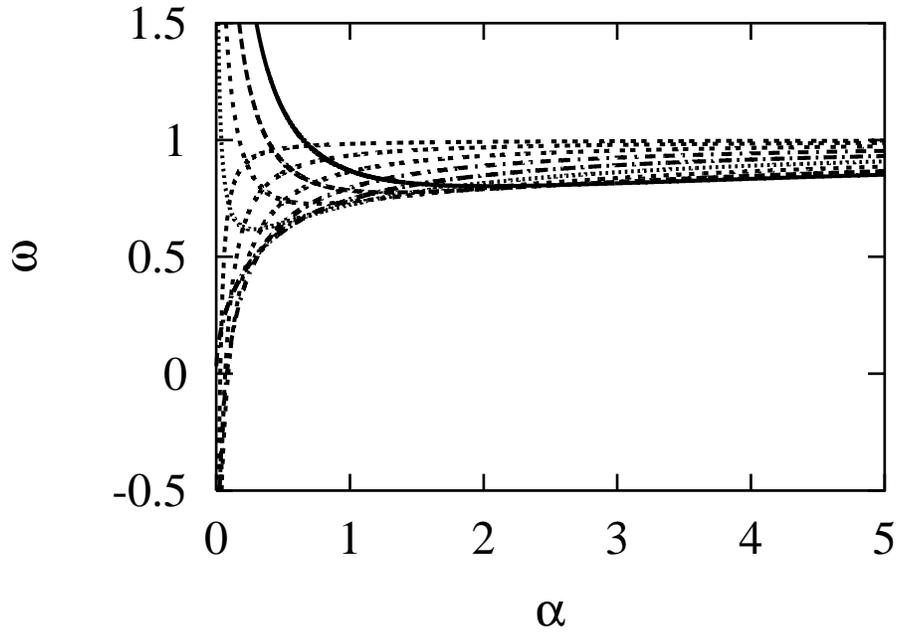}%
\caption{Weight factor as a function of $\alpha$.
The model and lines are the same as in Fig. \ref{fig:mchal}.}
\label{fig:mchaw}
\end{center}
\end{figure}

\begin{figure}
\begin{center}
\epsfig{file=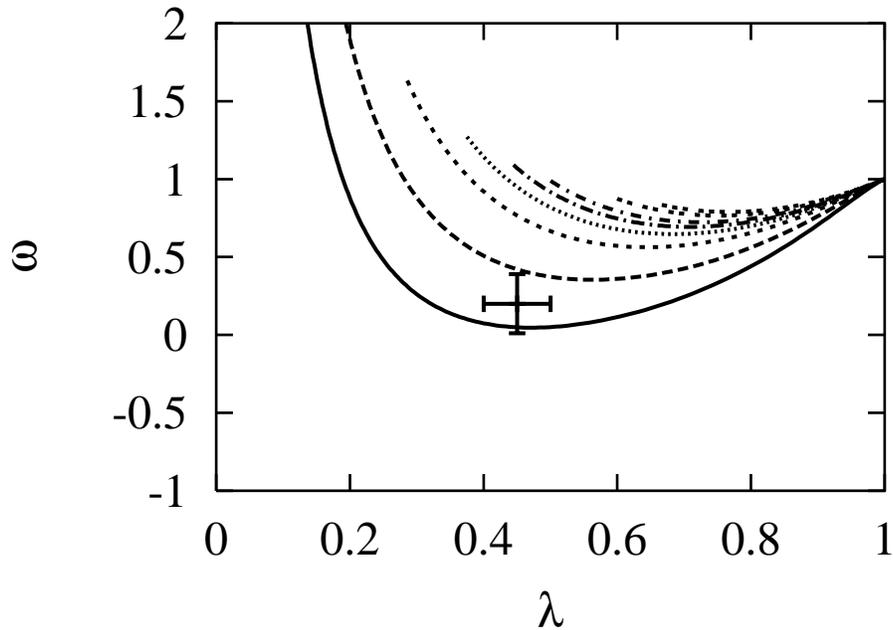}%
\caption{Weight factors as functions of chaoticities
in the model of multi-coherent sources and one chaotic source, 
varying $\epsilon$
from 0 to 1.
The lines from down to up stand for $\alpha=0.1$, 0.2, 0.4, 0.6, 0.8, 1.0,
1.5, 2.0, respectively.
The plot with error-bars
is the experimental data
from NA44.
}
\label{fig:mchalw}
\end{center}
\end{figure}

The chaoticity and weight factor are obtained as
\BEA
\lambda&=&\frac{\alpha}{\alpha+(1-\epsilon)^2},\\
\omega&=&
\frac{2\alpha^2+2\alpha(1-\epsilon)^2+3(1-\epsilon)^3(1-2\epsilon)}
{2\left(\alpha^2+3\alpha(1-\epsilon)^2+(1-\epsilon)^3\right)}
\sqrt{\frac{\alpha+(1-\epsilon)^2}{\alpha}}.
\EEA
Figures \ref{fig:mchal} and \ref{fig:mchaw} show $\lambda$ and $\omega$,
respectively,
as functions of $\alpha$ at $\epsilon=0.1-0.9$.
In Fig. \ref{fig:mchalw}, the weight factor is shown
as a function of the chaoticity, varying $\epsilon$ for various $\alpha$'s.
The divergence of $\omega$ at $\alpha=0$, except for $\epsilon=0.5$,
is caused by the incomplete cancellation of the $C_2$'s.

\section{Discussions and Summary}
\label{sec:sum}
\label{sec3}

We have examined various models of the source that are not completely
chaotic.   The chaoticity comes out to be between 0 and 1 in all models, 
but the weight factor takes a wide range of the value.  The value of the
weight factor even diverges in some cases, as a consequence of incomplete 
removal of the two-pion correlations from $C_3$.
For a partially coherent source that is commonly examined, the removal
is complete and the anomalous behavior does not appear.

There is a way to avoid such incomplete removal, at least in all the models
that we have examined here.  It is done through a new subtraction,
\BE
R_3(p_1,p_2,p_3)=C_3(p_1,p_2,p_3)-1-\frac{\langle n \rangle
  \left(2\langle n(n-1) \rangle - \langle n \rangle^2 \right)}
  { \langle n(n-1)(n-2) \rangle}
 \frac{ \langle n(n-1) \rangle}{\langle n \rangle^2}
 \sum_{(a,b)}\left(C_2(p_a,p_b)-1\right),
\label{eq:newr3}
\EE
where $\langle n \rangle$ and the similar are defined in Eqs. (\ref{eq:nave})
- (\ref{eq:naveend}).
This subtraction works correctly for the multi-coherent sources and also 
for the mixture of multi-coherent sources and one chaotic source 
(but at the zero relative momenta). 
We find that the terms corresponding to the ``other two-pion
correlations'' also vanish at the zero relative momenta (see Appendix B).
Using Eq. (\ref{eq:newr3}), we define a new weight factor as
\BE
\omega' = \frac{\langle n(n-1)(n-2) \rangle}{\langle n \rangle^3}
\frac{1}{2}R_3(p_1,p_2,p_3)
 \left(\frac{\langle n (n-1) \rangle}{\langle n \rangle^2}
\sum_{(a,b)}\left\{C_2(p_a,p_b)-1\right\}\right)^{-\frac{3}{2}}.
\label{eq:wfprime}
\label{EQ:WFPRIME}
\EE
For the partially coherent source, the new weight factor becomes the
conventional weight factor of Eq. (\ref{eq:wf}).

Though the new weight factor properly represents the strength of the genuine
three-pion correlations for all models examined in this work, the new weight
factor must not have this property at all times.  The relation between $C_2$'s
and $C_3$ is generally complicated, depending on dynamics in each case.
We thus do not expect that the new weight factor has a universal application.

\begin{figure}
\begin{center}
\epsfig{file=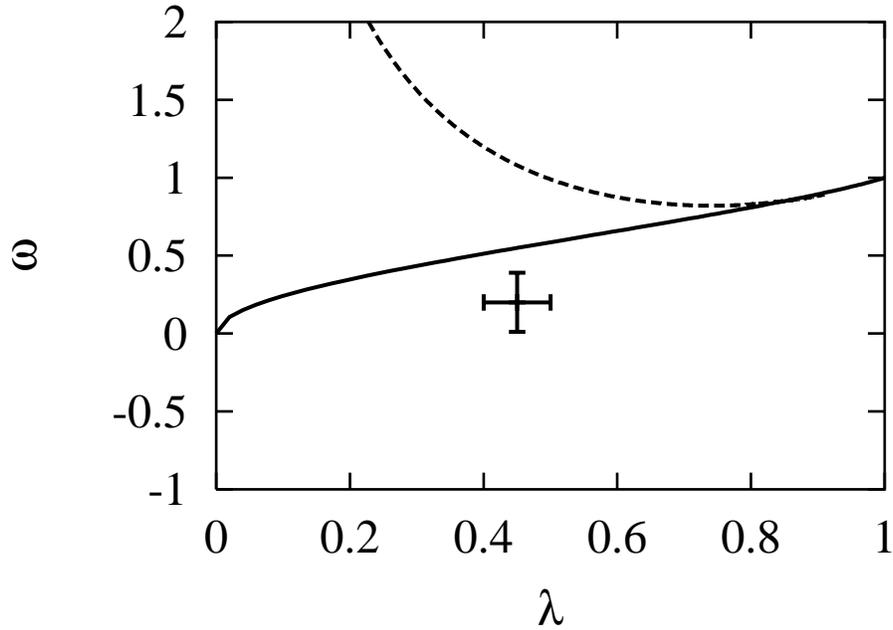}%
\caption{Weight factors as functions of chaoticities.
Solid and dashed lines stand for
the partially coherent source 
and the multi-coherent sources,
respectively. The plot with error-bars
is the experimental data
from NA44.
}
\label{fig:nondcc}
\end{center}
\end{figure}

Figures \ref{fig:nondcc} and \ref{fig:mchalw} illustrate the weight factors
as functions of the chaoticities for the different models, together with
the experimental data from the CERN NA44 Collaboration\cite{na44}.
In this experiment, the chaoticity and weight factor are
measured as $0.4-0.5$ and $0.20\pm0.19$, respectively.
Figure \ref{fig:nondcc} shows that the partially coherent source and
the multi-coherent sources disagree with the experiment.
The mixture of multi-coherent sources and one chaotic source 
reproduces the data if we set $\alpha = 0.13$ and $\epsilon=0.60$.
This corresponds to the mean number of the coherent sources being 0.13 and
to about 60 \% of the total pions emitted from the chaotic source.
The ratio of the pion number emitted from one 
coherent source to that from one chaotic source, $h_{11}/F_{11}$,
is then about 5.  This ratio may be unrealistically large, but we note that
the experimental data are ``minimum-bias,'' suggesting that our best-fit
may not be unrealistic since the multiplicity fluctuation can be large.
Further data are needed to confirm that this is indeed the case.

In summary,
we investigate the two- and three-pion correlations for various models
of a source that is not completely chaotic.
The chaoticity and weight factor are evaluated as measures
of two- and three-pion correlations. 
The chaoticity always varies between 0 and 1, but the weight factor takes
the value of a wide range and sometimes even diverges.  The conventional
weight factor includes the effects of the two-pion correlations in some models,
yielding the anomalous behavior.  We propose, in all model considered here,
a new weight factor that has
no such difficulty, but not expected to
be valid universally.  We find that the model of multi-coherent sources
and one chaotic source could reproduce
the chaoticity and the weight factor observed in the recent experiment.

\acknowledgements
We acknowledge informative and stimulating discussions with T. Humanic,
especially regarding the NA44 experiment.
This research is supported
by the U.S.~Department of Energy under grant DE-FG03-87ER40347 at CSUN,
and the U.S.~National Science
Foundation under grants PHY88-17296 and PHY90-13248 at Caltech.

\vfill\eject

\appendix
\section{multi-coherent source model}
\label{appA}

We assume that $N$ coherent sources are created during a collision
and that the position of the $n$-th source is $X_n$.
The sources are uncorrelated with each other.
A source current  is defined as
\BE
J(x) = \sum_{n=1}^N j(x-X_n) e^{-i \theta_n}.
\EE
where $\theta_n$ is a unique random number varying form $0$ to $2\pi$.
The average about $\theta_n$ is denoted by $\langle \cdots \rangle$.
The Fourier transform of the source current is written as
\BE
J(p) = \sum_{n=1}^N j(p) e^{ i p\cdot X_n -i\theta_n},
\EE
where $j(p)$ is the Fourier transform of $j(x)$.
The one-pion spectrum is  
\BEA
W_N(p_1) &=& \frac{1}{(2\pi)^3 p^0_1}\langle |J(p_1)|^2 \rangle \nonumber \\
&=& \frac{|j(p_1)|^2}{(2\pi)^3 p_1^0} \sum^N_{n,m} e^{-ip_1\cdot(X_n-X_m)}
\langle e^{i \theta_n - i \theta_m}\rangle \nonumber \\
&=& a_1 N,
\EEA
where $p^0$ is on-shell and
\BE
a_1 = \frac{|j(p_1)|^2}{(2\pi)^3 p^0_1}.
\EE
The two-pion spectrum is 
\BEA
W_N(p_1,p_2) &=& \frac{1}{(2\pi)^6 p^0_1p_2^0} 
  \langle |J(p_1)|^2 |J(p_2)|^2 \rangle \nonumber \\
&=& a_1a_2  \left[N^2 + \sum^N_{n\neq m} e^{-i(p_1-p_2)\cdot(X_m-X_n)}
\right]
\EEA
The location of each small source, $X_n$, is assumed to obey a distribution
$\rho(X_n)$, which is normalized to be unity, or $\int\rho(X_n)dX_n = 1$.
Averaging the two-pion spectrum with the distribution,
we obtain
\BEA
\overline{W}_N(p_1,p_2)&=&\int\rho(X_1)dX_1 \cdots \rho(X_N)dX_N
W_N(p_1,p_2)
 \nonumber \\
&=& a_1a_2 \left[N^2+N(N-1)|\rho_{12}|^2\right],
\EEA
where $\rho_{12} = \int \rho(x)dx e^{-i(p_1-p_2)\cdot x}$.
Note that $\sum_{i,j} A_{ij} = \sum_i A_{ii}+\sum_{i\neq j}A_{ij}$
and
$\sum_{i\neq j} 1 = N(N-1)$.

The three-pion spectrum is
\BEA
W_N(p_1,p_2,p_3) &=&  \frac{1}{(2\pi)^6 p^0_1p_2^0p_3^0}
  \langle |J(p_1)|^2 |J(p_2)|^2|J(p_3)|^2 \rangle,
 \nonumber \\
\EEA
and the averaged spectrum is evaluated as
\BEA
\overline{W}_N(p_1,p_2,p_3)
&=& a_1a_2a_3\left[
  N^3+N^2(N-1)\left(|\rho_{12}|^2+|\rho_{23}|^2+|\rho_{31}|^2\right)
\right.\nonumber \\ && \left.
  +N(N-1)(N-2)\cdot 2{\mathrm{Re}}\left(\rho_{12}\rho_{23}\rho_{31}\right)
\right].
\EEA
Note that $\sum_{i,j,k}A_{ijk} = \sum_i A_{iii} 
+ \sum_{i\neq j}\left(A_{iij}+A_{iji}+A_{jii}\right)
+\sum_{i\neq j \neq k} A_{ijk}$ and that $\sum_{i\neq j \neq k}1=N(N-1)(N-2)$.

We next assume that the coherent sources obey the Poisson distribution.
The Poisson distribution is 
\BE
\label{eq:usupoi}
P_N^{\rm{(usual)}} = \frac{\alpha^N}{N!}e^{-\alpha}
\quad \mbox{for } N=0\sim\infty.
\EE
The $N=0$  event should be excluded because in such an event no pions are
emitted. The distribution we use is renormalized as
\BE
\label{eq:poisson}
P_N = \frac{\alpha^N}{N!}\frac{1}{e^\alpha-1}
\quad \mbox{for } N=1\sim\infty.
\EE
The expectation values change from those of the usual Poisson distribution,
\BEA
\langle N \rangle_P &=& \frac{\alpha}{1-e^{-\alpha}}, \\
\langle N(N-1) \rangle_P &=& \frac{\alpha^2}{1-e^{-\alpha}}, \\
\langle N(N-1)\cdots(N-n)\rangle_P &=& \frac{\alpha^{n+1}}{1-e^{-\alpha}}. 
\EEA

Pion spectra averaged with the Poisson distribution, 
$W = \sum_{N=1}^\infty P_N\overline{W}_N$,
are
\BEA
\label{eq:aw1}
W(p_1) &=& a_1 \frac{\alpha}{1-e^{-\alpha}},\\
W(p_1,p_2) &=& \frac{a_1a_2}{1-e^{-\alpha}}
\left[\alpha(\alpha+1)+\alpha^2|\rho_{12}|^2 \right],\\
W(p_1,p_2,p_3) &=&  \frac{a_1a_2a_3}{1-e^{-\alpha}}
\left[
  \alpha(\alpha^2+3\alpha+1)
\right. \nonumber \\ && \left.
  +\alpha^2(\alpha+2)\left(|\rho_{12}|^2+|\rho_{23}|^2+|\rho_{31}|^2\right)
  +\alpha^3\cdot 2{\mathrm{Re}}\left(\rho_{12}\rho_{23}\rho_{31}\right)
\right].
\label{eq:aw3}
\EEA
The correlation functions are
\BEA
{C}_2(p_1,p_2) &=& 1+\frac{\alpha}{\alpha+1} |\rho_{12}|^2,\\
{C}_3(p_1,p_2,p_3) &=& 1+\frac{\alpha(\alpha+2)}{\alpha^2+3\alpha+1}
  \left(|\rho_{12}|^2+|\rho_{23}|^2+|\rho_{31}|^2\right)
\nonumber \\ &&
 + \frac{\alpha^2}{\alpha^2+3\alpha+1}
     \cdot 2{\mathrm{Re}}\left(\rho_{12}\rho_{23}\rho_{31}\right),
\EEA
where the correlation functions are normalized to be unity at $|p_i-p_j|\to
\infty$. The chaoticity and weight factor are
\BEA
{\lambda} &=& \frac{\alpha}{\alpha+1},\\
{\omega} &=& \frac{1}{2}\frac{2\alpha^2+2\alpha+3}{\alpha^2+3\alpha+1}
\sqrt{\frac{\alpha+1}{\alpha}}.
\EEA

\section{Derivation of Eq. (\ref{eq:wfprime})}
\label{appB}

The two- and three-pion spectra can be written generally as
\BEA
W_2(p_1,p_2) &=&\langle a_1^\dagger a_1 \rangle \langle a_2^\dagger a_2 \rangle
+\langle a_1^\dagger a_2 \rangle \langle a_2^\dagger a_1 \rangle
+\llangle  a_1^\dagger a_2^\dagger a_1 a_2 \rrangle,
\\
W_3(p_1,p_2,p_3) &=&
\langle a_1^\dagger a_1 \rangle \langle a_2^\dagger a_2 \rangle
 \langle a_3^\dagger a_3 \rangle
+\sum_{(a,b,c)}\langle a_a^\dagger a_a \rangle 
\langle a_b^\dagger a_c \rangle \langle a_c^\dagger a_b \rangle
+2{\mathrm{Re}}\left(\langle a_1^\dagger a_2 \rangle 
\langle a_2^\dagger a_3 \rangle \langle a_3^\dagger a_1 \rangle\right)
\nonumber \\ &&
+\sum_{(a,b,c)}\langle a_a^\dagger a_a \rangle
\llangle a_b^\dagger a_c^\dagger a_b a_c \rrangle
+ 2{\mathrm{Re}} \sum_{(a,b,c)}\langle a_a^\dagger a_b \rangle
\llangle a_b^\dagger a_c^\dagger a_a a_c \rrangle
+\llangle  a_1^\dagger a_2^\dagger  a_3^\dagger a_1 a_2 a_3 \rrangle,
\nonumber \\
\label{eq:w3br}
\EEA
where $a_i$ is an annihilation operator of momentum $p_i$.
We assume that $\langle a_i \rangle = 0$, and therefore we don't consider
the coherent source and partially coherent source in this appendix.
$ \llangle\cdots\rrangle$ corresponds to  a cumulant, 
or a connected Green function.
If we have a generating functional, $Z[z^*(p),z(p)]$,  such as
\BE
 \langle a_1^\dagger\cdots a_n^\dagger a_1\cdots a_n \rangle
=\left.
\frac{\delta^{2n} Z[z^*(p),z(p)]}{\delta z(p_1)\cdots\delta z(p_n)
 \delta z^*(p_1)\cdots\delta z(p_n)}\right|_{z=0},
\EE
then the cumulant is obtained as
\BE
 \llangle a_1^\dagger\cdots a_n^\dagger a_1\cdots a_n \rrangle
=\left.
\frac{\delta^{2n}\left(\ln Z[z^*(p),z(p)]\right)}
{\delta z(p_1)\cdots\delta z(p_n)
 \delta z^*(p_1)\cdots\delta z(p_n)}\right|_{z=0}.
\EE

We introduce two assumptions.
The first is   
\BEA
\llangle  a_1^\dagger a_2^\dagger a_1 a_2 \rrangle
&=& A \langle a_1^\dagger a_1 \rangle \langle a_2^\dagger a_2 \rangle \\
\llangle  a_1^\dagger a_2^\dagger a_3^\dagger a_1 a_2 a_3\rrangle
&=& B \langle a_1^\dagger a_1 \rangle \langle a_2^\dagger a_2 \rangle
  \langle a_3^\dagger a_3 \rangle,
\EEA
where $A$ and $B$ are constant.
The second assumption is that
the interference terms, such as 
$ \langle a_1^\dagger a_2 \rangle \langle a_2^\dagger a_1 \rangle$,
vanish by integrating over the momenta, for example,
\BE
\int d^3p_1d^3p_2  
\langle a_1^\dagger a_2 \rangle \langle a_2^\dagger a_1 \rangle
=0.
\EE
Thus we can obtain
\BEA
\langle n(n-1) \rangle &=& (1+A)\langle n \rangle^2, \\
\langle n(n-1)(n-2) \rangle &=&
(1+3A+B)\langle n \rangle^3.
\EEA
Correlation functions at the zero relative momenta become
\BEA
C_2(p,p) &=& 1+\frac{1}{1+A},\\
C_3(p,p,p) &=& 1+\frac{1}{1+3A+B}
\left\{ 3 + 2 + 6A \right\}.
\EEA
The terms 3, 2 and $6A$ between the braces in the above equation
correspond to $\sum_{(a,b,c)}\langle a_a^\dagger a_a \rangle
\langle a_b^\dagger a_c \rangle \langle a_c^\dagger a_b \rangle $,
$2{\mathrm{Re}}\left(\langle a_1^\dagger a_2 \rangle
\langle a_2^\dagger a_3 \rangle \langle a_3^\dagger a_1 \rangle\right)$
and $2{\mathrm{Re}} \sum_{(a,b,c)}\langle a_a^\dagger a_b \rangle
\llangle a_b^\dagger a_c^\dagger a_a a_c \rrangle$ in Eq. (\ref{eq:w3br}).
To obtain the genuine three-pion correlation,
we should subtract the first and third terms between the braces.
Thus we define the new weight factor, considering the above discussion, as
\BEA
\omega'&=&\frac{1}{2}\frac{
(1+3A+B)\left(C_3(p,p,p)-1\right)- 3(1+2A) \left(C_2(p,p)-1\right) }
{\left[(1+A)\left(C_2(p,p)-1\right)\right]^{\frac{3}{2}}}
\nonumber \\
&=&\frac{1}{2}\frac{
\frac{\langle n(n-1)(n-2) \rangle}{\langle n \rangle^3}
\left(C_3(p,p,p)-1\right)
- 3\cdot
\frac{2\langle n (n-1) \rangle -\langle n \rangle^2}{\langle n \rangle^2}
 \left(C_2(p,p)-1\right) }
{\left[\frac{\langle n (n-1) \rangle}{\langle n \rangle^2}
\left(C_2(p,p)-1\right)\right]^{\frac{3}{2}}}
\label{eq:newwf}
\EEA
This equation corresponds to Eq. (\ref{eq:wfprime}).

The two assumptions are valid for multi-coherent sources 
and one chaotic source with setting $A= \alpha^{-1}(1-\epsilon)^2$ 
and $B=\alpha^{-2}(1-\epsilon)^3$.
For multi-coherent sources, the assumptions are not valid because of
the factor $(1-\exp(-\alpha))^{-1}$, in Eqs. (\ref{eq:aw1})-(\ref{eq:aw3}),
due to the renormalized Poisson distribution, Eq. (\ref{eq:poisson}).
In using the usual Poisson distribution instead of the renormalized one,
the factor vanishes, and the assumptions become applicable
with $A=\alpha^{-1}$ and $B=\alpha^{-2}$. The correlation functions
are the same both in the cases of the usual and renormalized distributions.
If we use the renormalized distributions and other assumptions,
we can derive Eq. (\ref{eq:newwf}), but the derivation is more complicated.


\end{document}